\documentclass[12pt]{article}
\usepackage[dvips]{graphicx}
\usepackage{amsmath,amssymb}
\usepackage{cases}
\usepackage{bm}
\usepackage{cite}

\title{A viscous solution of the spherical vortex to the Navier-Stokes equations}
\author{}

\begin{document}
\rightline{December 2014}
\rightline{~~~~~~~~}
\vskip 1cm
\centerline{\large \bf A viscous solution of the spherical vortex}
\centerline{\large \bf to the Navier-Stokes equations}
\vskip 1cm

\centerline
{{Minoru Fujimoto${}^1$
 }, 
 {Kunihiko Uehara${}^2$
 } and 
 {Shinichiro Yanase${}^3$
 }
}
\vskip 1cm
\centerline{\it ${}^1$Seika Science Research Laboratory,
Seika-cho, Kyoto 619-0237, Japan}
\centerline{\it ${}^2$Department of Physics, Tezukayama University,
Nara 631-8501, Japan}
\centerline{\it ${}^3$Graduate School of Natural Science and Technology, Okayama University,
Okayama 700-8530, Japan}
\vskip 1cm
\centerline{\bf Abstract}
\vskip 0.2in
  We deal with the Hill's spherical vortex, which is 
an exact solution to the Euler equation, 
and manage the solution to satisfy the incompressible Navier-Stokes(INS) 
equations with a viscous term. 
Once we get a viscous solution to the INS equations, 
we will be able to analyze the flows with discontinuities in vorticity. 

  In the same procedure, we also present a time developing exact solution 
to the INS equations, which has a rotation on the axis 
besides the Hill's vortex. 
\vskip 5mm
%
\noindent
PACS number(s): 47.10.ad, 47.15.ki, 47.32.-y
\vskip 5mm

\setcounter{equation}{0}
\addtocounter{section}{0}
\hspace{\parindent}

  It is a well-known fact for a century 
that the Hill's spherical vortex\cite{Hill} is one of 
exact solutions\cite{OBrien,Terrill,Wang,Weinbaum} to the Euler equations. 
A characteristic feature of the vortex is that 
the vorticity only exists inside the sphere, 
where the vorticity is referred using by the velocity $\boldsymbol{u}$ as
\begin{equation}
  \boldsymbol{\omega}=\boldsymbol{\nabla}\times\boldsymbol{u}.
\end{equation}
The velocity inside or outside of the sphere radius $a$ is given by 
\begin{equation}
\begin{cases}
  \boldsymbol{u}_\text{inside}=
  \begin{pmatrix}
    u_\eta(t,\eta,\varphi,z)\\ 
    u_\varphi(t,\eta,\varphi,z)\\ 
    u_z(t,\eta,\varphi,z)
  \end{pmatrix}=
  \begin{pmatrix}
    \displaystyle{\frac{3U}{2a^2}\eta z}\\ 
    0\\ 
    \displaystyle{\frac{3U}{2a^2}(a^2-2\eta^2-z^2)}
  \end{pmatrix},\\
  \vspace{-4mm}\\
  \boldsymbol{u}_\text{outside}=
  \begin{pmatrix}
    u_\eta(t,\eta,\varphi,z)\\ 
    u_\varphi(t,\eta,\varphi,z)\\ 
    u_z(t,\eta,\varphi,z)
  \end{pmatrix}=
  \begin{pmatrix}
    \displaystyle{\frac{3a^3U}{2}\frac{\eta z}{(\eta^2+z^2)^{5/2}}}\\
    0\\ 
    \displaystyle{-U-\frac{a^3U}{2}\frac{(\eta^2-2z^2)}{(\eta^2+z^2)^{5/2}}}
  \end{pmatrix},\\
\end{cases}
\end{equation}
where we take the cylindrical coordinates $(\eta,\varphi,z)$ 
and $-U$ is a value of the velocity for the fluid at $z\to\pm\infty$.
The continuity condition is satisfied with each velocity. 

  We manage the solution above to satisfy the incompressible Navier-Stokes(INS) 
equations with the viscous term,
\begin{equation}
\frac{\partial \boldsymbol{u}}{\partial t}
    +(\boldsymbol{u}\cdot\nabla)\boldsymbol{u}
  =\boldsymbol{K}-\frac{1}{\rho}\nabla p
    +\nu\triangle\boldsymbol{u},
\end{equation}
where $\boldsymbol{K}$ is an external force vector,
$p$ is the pressure and $\rho$ is the density
and $\nu$ is the viscosity.
The key to carry this idea out is the pressure term and the external force 
in the INS equations. 
  The condition for the pressure is the continuation 
between a pressure inside and outside at the sphere radius $a$. 
But we do not impose the continuity condition for the external force. 
We discuss the discontinuity for the external force afterward. 

  The results we have got for the pressures inside and outside are
\begin{equation}
  \begin{cases}
    p_\text{inside}(t,\eta,\varphi,z)=\displaystyle{-\frac{9U^2\rho}{8a^4}
                   \left\{-\eta^4+z^4+a^2(\eta^2-2z^2)\right\}},\\ 
    \vspace{-4mm}\\
    p_\text{outside}(t,\eta,\varphi,z)=\displaystyle{\frac{U^2\rho}{8}\left\{5-\displaystyle{\frac{4a^3(\eta^2-2z^2)}{(\eta^2+z^2)^{5/2}}
                  -\frac{a^6(\eta^2+4z^2)}{(\eta^2+z^2)^4}}\right\}},
  \end{cases}
\end{equation}
and for the external force inside and outside are
\begin{equation}
  \begin{cases}
  \boldsymbol{K}_\text{inside}=
  \begin{pmatrix}
    K_\eta(t,\eta,\varphi,z)\\ 
    K_\varphi(t,\eta,\varphi,z)\\ 
    K_z(t,\eta,\varphi,z)
  \end{pmatrix}=
  \begin{pmatrix}
    0\\ 
    0\\ 
    \displaystyle{\frac{15U}{a^2}\nu}
  \end{pmatrix},\\
    \boldsymbol{K}_\text{outside}(t,\eta,\varphi,z)=\boldsymbol{0}.
  \end{cases}
\end{equation}

  It is usually expected that the spatial discontinuity for the vorticity 
may take place in the turbulence, as in the case of the Hill's vortex. 
This means that a spatial differential of the velocity at the point 
would be disconnected and the Laplacian term in the INS equation 
would be divergent. 
When we get a viscous solution to the INS equations above, 
we can utilize the solution to analyze the flows 
with discontinuities in vorticity using a smoothing function, 
which tends to the step function as some parameter tends to zero. 

  The reason why we have not imposed the continuity condition 
for the external force is that 
we can set the step function for the external force in the experiments 
or the thought experiments 
which is difficult technically in actual experiments though. 

  In the same scenario we would like to present another set of solutions, 
which has a time developing rotation on the $z$-axis besides the Hill's vortex. 
The velocities inside and outside of the sphere radius $a$ are
\begin{equation}
\begin{cases}
  \boldsymbol{u}_\text{inside}=
  \begin{pmatrix}
    u_\eta(t,\eta,\varphi,z)\\ 
    u_\varphi(t,\eta,\varphi,z)\\ 
    u_z(t,\eta,\varphi,z)
  \end{pmatrix}=
  \begin{pmatrix}
    \displaystyle{\frac{3U}{2a^2}\eta z}\\ 
    \displaystyle{\frac{1}{\eta}T(t)}\\ 
    \displaystyle{\frac{3U}{2a^2}(a^2-2\eta^2-z^2)}
  \end{pmatrix},\\
  \vspace{-4mm}\\
  \boldsymbol{u}_\text{outside}=
  \begin{pmatrix}
    u_\eta(t,\eta,\varphi,z)\\ 
    u_\varphi(t,\eta,\varphi,z)\\ 
    u_z(t,\eta,\varphi,z)
  \end{pmatrix}=
  \begin{pmatrix}
    \displaystyle{\frac{3a^3U}{2}\frac{\eta z}{(\eta^2+z^2)^{5/2}}}\\
    \displaystyle{\frac{1}{\eta}T(t)}\\ 
    \displaystyle{-U-\frac{a^3U}{2}\frac{(\eta^2-2z^2)}{(\eta^2+z^2)^{5/2}}}
  \end{pmatrix},\\
\end{cases}
\end{equation}
where $T(t)$ is any function of $t$, 
which will be compensated by the pressure term given below. 
The pressures inside and outside of the sphere radius $a$ are
\begin{equation}
  \begin{cases}
    p_\text{inside}(t,\eta,\varphi,z)=\displaystyle{-\frac{\rho}{8a^4\eta^2}
                   \left\{9\eta^2U^2\left(-\eta^4+z^4+a^2(\eta^2-2z^2)\right)
                   +4a^4\left(T(t)^2+2\eta^2\varphi T'(t)\right)\right\}},\\ 
    \vspace{-4mm}\\
    p_\text{outside}(t,\eta,\varphi,z)=\rho\left[\displaystyle{\frac{U^2}{8}\left\{5-\displaystyle{\frac{4a^3(\eta^2-2z^2)}{(\eta^2+z^2)^{5/2}}
                  -\frac{a^6(\eta^2+4z^2)}{(\eta^2+z^2)^4}}\right\}
                  +\frac{T(t)^2}{2\eta^2}+\varphi T'(t)}\right],
  \end{cases}
\end{equation}
and for the external force inside and outside are same as before as
\begin{equation}
  \begin{cases}
  \boldsymbol{K}_\text{inside}=
  \begin{pmatrix}
    K_\eta(t,\eta,\varphi,z)\\ 
    K_\varphi(t,\eta,\varphi,z)\\ 
    K_z(t,\eta,\varphi,z)
  \end{pmatrix}=
  \begin{pmatrix}
    0\\ 
    0\\ 
    \displaystyle{\frac{15U}{a^2}\nu}
  \end{pmatrix},\\
    \boldsymbol{K}_\text{outside}(t,\eta,\varphi,z)=\boldsymbol{0}.
  \end{cases}
\end{equation}
This set of solutions looks like a coordinate transformation to 
a rotating coordinate, but this is not true because of the existence 
of the denominator $\eta$. 
We will discuss a solution of the rotating vortex only within a sphere 
in a separate paper. 
There exists singularity in $z$-axis for this solution though, 
some singular type of solutions would be used to analyze for the case of 
tangled vortex lines\cite{Moffatt}. 

\vskip 5mm

\end{document}